\ifdefined\XeTeXrevision\else\pdfoutput=1\fi 
\documentclass[11pt]{article}
\usepackage{arxiv}
\usepackage[utf8]{inputenc}
\usepackage[T1]{fontenc}
\usepackage{amsmath,amssymb,amsthm}
\usepackage{graphicx}
\usepackage{booktabs}
\usepackage[hidelinks]{hyperref}

\theoremstyle{plain}
\newtheorem{theorem}{Theorem}

\newtheorem{assumption}{Assumption}
\theoremstyle{definition}
\newtheorem{definition}{Definition}
\theoremstyle{remark}
\newtheorem{remark}{Remark}

\renewenvironment{keywords}
  {\par\addvspace\medskipamount\noindent\textbf{\textit{Keywords:}}\enspace\ignorespaces}
  {\par}

\newcommand{\R}{\mathbb{R}}
\newcommand{\Rp}{\mathbb{R}_{\ge 0}}
\newcommand{\sh}{\hat s}
\newcommand{\lm}{\lambda_{m}}
\newcommand{\lM}{\lambda_{M}}
\newcommand{\eb}{\bar e}
\newcommand{\kb}{\bar\kappa}
\newcommand{\norm}[1]{\lVert #1\rVert}
\newcommand{\dd}{\,\mathrm{d}}

\renewcommand{\shorttitle}{Velocity-Free Funnel Control of Euler--Lagrange Systems}
\renewcommand{\headeright}{Preprint}
\renewcommand{\undertitle}{Preprint}

\title{Velocity-Free Funnel Control of Euler--Lagrange Systems: An~Artificial-Delay Approach%
\thanks{This work has been submitted to the European Control Conference (ECC) 2027.
This work was supported by ANRF ARG Grant ANRF/ARG/2025/003876/ENS.}}

\author{Chayan Kumar Paul\\
  Department of Cyber-Physical Systems\\
  Indian Institute of Science, Bengaluru 560012, India\\
  \texttt{chayanpaul007@gmail.com}
  \And
  Pushpak Jagtap\\
  Department of Cyber-Physical Systems\\
  Indian Institute of Science, Bengaluru 560012, India\\
  \texttt{pushpak@iisc.ac.in}}

\date{}

\begin{document}
\maketitle

\begin{abstract}
Classical funnel control keeps the tracking error within a prescribed, time-varying envelope by increasing the feedback gain as the error approaches the boundary. For mechanical systems, this guarantee typically relies on velocity measurements. We show that it can be retained for fully actuated Euler--Lagrange systems using only position measurements. The proposed controller replaces the velocity error with a difference quotient of the position error over a fixed delay, while preserving the classical unbounded gain. It requires no plant model, reference derivatives, observer, or filter states. The central challenge is to reconcile the averaging inherent in delayed feedback with a gain that can grow without bound near the funnel boundary. We resolve this by establishing an a priori bound on the gain along closed-loop trajectories, independent of the delay. This bound yields a computable delay threshold that guarantees tracking within the prescribed funnel and boundedness of all closed-loop signals. The threshold depends only on bounds on the inertia, forces, and reference. Simulations demonstrate tracking accuracy comparable to full-state funnel control and tolerance of delays substantially longer than certified. With quantized position measurements, the controller also requires substantially less torque than an existing derivative-free funnel controller.
\end{abstract}


\section{Introduction}
Funnel control~\cite{Ilchmann2002} keeps the tracking error of an uncertain system within a time-varying funnel chosen by the designer. The controller gain is neither identified nor tuned: it increases as the error approaches the funnel boundary, so that the transient behavior is prescribed rather than merely bounded. The controller is model-free, in the sense that it relies only on structural properties, namely a known relative degree, a sign-definite high-frequency gain, and stable internal dynamics~\cite{Berger2018,Berger2021,Berger2025survey}.  Applications to servo drives and robot manipulators demonstrate the relevance of this approach to motion control~\cite{Hackl2013,Hackl2017,HacklKennel2012}.

For fully actuated manipulators, the joint positions have relative degree two with respect to the applied torques. Funnel controllers for this class therefore use the velocity error, either directly or in combination with the position error~\cite{Hackl2013,Berger2018,Berger2021,Dennstaedt2025low}. Joint encoders, however, measure only position. Tracking from position measurements alone is a classical problem in the control of Euler--Lagrange (EL) systems~\cite{Nicosia1990,Berghuis1993,Kelly2005}. Approximate (``dirty'') derivatives, for example, yield asymptotic tracking~\cite{Loria2016}, but this guarantee does not prescribe how the error evolves during the transient.

Several approaches address this additional requirement. High-gain observers estimate the missing derivatives~\cite{Chowdhury2019,Dimanidis2020,KhalilPraly2014}, although choosing a sufficiently large observer gain is difficult without an explicit threshold. Dynamic compensators avoid derivative estimates at the cost of additional controller states. Examples include backstepping-type filters~\cite{Ilchmann2007}, funnel pre-compensators~\cite{BergerReis2018tac,Lanza2022}, and input filters~\cite{Dennstaedt2025filter,Schaa2026}. The input filters also require a constant input matrix, which excludes manipulators with configuration-dependent inertia. A third approach uses finite differences. The closest work to ours~\cite{Lanza2024derivfree} uses two consecutive position samples and establishes prescribed transient bounds for the sampled controller, including under configuration-dependent inertia. Worst-case bounds on the dynamics determine the sampling interval and a minimum gain, while normalization keeps the input bounded. Here, we instead form a difference quotient from current and delayed position measurements and retain the classical, unbounded funnel gain.

Using a delay to recover derivative information connects this construction to \emph{artificial-delay} feedback~\cite{FridmanSelivanov2025}. The underlying principle is that, for suitable system classes, derivative feedback can be replaced by delayed output feedback when the delay is sufficiently small~\cite{IlchmannSangwin2004,French2009}. This principle underlies proportional(-integral)-retarded controllers~\cite{Ramirez2016}, with applications to servo drives and manipulators~\cite{Villafuerte2013,RamirezNeria2019}. Lyapunov--Krasovskii conditions are available for vector second-order systems~\cite{FridmanShaikhet2016}, sampled-data implementations~\cite{Selivanov2018}, and static sliding-mode control~\cite{Seuret2009,Xu2020}. Delayed measurements are also used in robotics to compensate unknown dynamics~\cite{YoucefToumi1990,Roy2021}. These results, however, do not establish tracking within prescribed time-varying bounds. Existing work on delays in the funnel and prescribed performance control typically treats them as impediments to this objective, adjusting the performance bounds to accommodate their effects~\cite{Bikas2023,Berger2025delays}. 

Our aim is to use the delay deliberately to remove velocity measurements while preserving the prescribed tracking guarantee. The controller estimates the velocity error by comparing the current position error with an earlier value. This avoids the need for an observer or filter, but the estimate reflects motion over a short interval. As a result, a change in torque takes time to register fully in the feedback signal. This matters near the funnel boundary, where the controller may sharply increase its gain to keep the tracking error within the funnel. To choose the delay before running the controller, we need to know how large the gain can become. For fully actuated EL systems, we show that this can be determined in advance without first choosing the delay. We then use this information to identify a range of delays for which the tracking error is guaranteed to stay inside the prescribed funnel. Sampled-data funnel control~\cite{LanzaSD2024,Lanza2024derivfree} selects the sampling period similarly, but bounds the input by normalizing the control law, with a gain above a threshold set by worst-case bounds on the dynamics, and relies on sampled velocities~\cite{LanzaSD2024} or a zero initial tracking error~\cite{Lanza2024derivfree}. We instead retain the classical continuous-time law, whose gain bound follows from the closed loop, so that no gain threshold is needed and the delay is the only additional parameter. 

The main contribution of the paper is threefold.
\begin{enumerate}
  \item A funnel controller for fully actuated EL systems that uses only position measurements. It obtains velocity information from delayed position measurements and requires neither an observer nor a model of the system.
  \item A proof that this controller, with the classical funnel gain and no saturation, keeps the tracking error within the prescribed funnel and all closed-loop signals bounded in continuous time whenever the delay is small enough.
  \item An explicit bound on the admissible delay, which shows that the delay is limited by the inertia of the system relative to the forces acting on it.
\end{enumerate}
The results are illustrated on a geared two-link arm, where the controller is compared with full-state funnel control, observer- and filter-based velocity estimates, proportional-retarded control and the sample-and-hold controller of~\cite{Lanza2024derivfree}.

\emph{Notation.} $\norm{\cdot}$ denotes the Euclidean norm and the induced matrix norm; $A\succeq B$ means that $A-B$ is positive semidefinite; $\|f\|_\infty$ is the essential supremum of $\norm{f(\cdot)}$; $W^{k,\infty}(I;\R^n)$ is the space of functions on $I$ which are essentially bounded together with their first $k$ (weak) derivatives.

\section{Problem Formulation}
\label{sec:problem}
Consider a fully actuated mechanical system with $n$ degrees of freedom,
\begin{equation}
M(q)\ddot q + C(q,\dot q)\dot q + g(q) + F(\dot q) + d(t) = \tau ,
\label{eq:EL}
\end{equation}
where $q(t)\in\R^n$ and $\tau(t)\in\R^n$ are the joint positions and torques, $M(q)$ is the inertia matrix, $C(q,\dot q)\dot q$ collects Coriolis and centrifugal terms, $g(q)$ is the gravity torque, $F(\dot q)$ the friction torque and $d(t)$ an external disturbance~\cite{Kelly2005}. Let $N(q,\dot q):=C(q,\dot q)\dot q+g(q)+F(\dot q)$.

\begin{assumption}\label{ass:sys}
\begin{enumerate}
  \item[(a)] $M\in C^1(\R^n;\R^{n\times n})$ is symmetric and there are known constants $0<\lm\le\lM$ such that $\lm I\preceq M(q)\preceq\lM I$ for all $q\in\R^n$.
  \item[(b)] $N$ is locally Lipschitz and $d\in L^\infty(\Rp;\R^n)$. Neither is used by the
  controller; bounds on compact sets enter only the admissible delay.
  \item[(c)] Only $q$ is measured, and its values over the last $h$ seconds can be stored, where $h>0$ is a design parameter.
  \item[(d)] $q_d\in W^{2,\infty}([-h,\infty);\R^n)$ can be evaluated at the current and past times; its derivatives are not used by the controller.
  \item[(e)] $q(\theta)=q^0$ for all $\theta\in[-h,0]$ and $\dot q(0)=0$.
\end{enumerate}
\end{assumption}

\begin{remark}
Assumption~\ref{ass:sys}(a) holds for manipulators with revolute joints~\cite{Kelly2005}; with an uncertain payload, the bounds are taken over the admissible payload range. Item~(b) covers viscous and smoothed Coulomb friction, and item~(e) describes a robot released from its brakes. Elastic joints, common with harmonic drives, are not covered; for them, the link position has a relative degree of 4.
\end{remark}

By~\eqref{eq:EL},
\begin{equation}
\ddot q = M(q)^{-1}\big(\tau - N(q,\dot q) - d(t)\big),
\label{eq:acc}
\end{equation}
that is, the position has relative degree two and the high-frequency gain $M(q)^{-1}$ is configuration dependent, with $\lM^{-1}I\preceq M(q)^{-1}\preceq\lm^{-1}I$.

\begin{definition}\label{def:funnel}
For $\alpha>0$, $\Psi_\alpha$ denotes the set of functions $\psi\in W^{1,\infty}(\Rp;\R)$ with $\inf_{t\ge0}\psi(t)>0$ and $\dot\psi(t)\ge-\alpha\psi(t)$ for almost all $t\ge0$.
\end{definition}

The funnel may thus shrink, but not faster than at the rate $\alpha$; a typical choice is $\psi(t)=(\psi_0-\psi_\infty)e^{-\alpha t}+\psi_\infty$. Let $e:=q-q_d$. The objective is a control law which uses only $q$ and $q_d$ and ensures, for a given $\psi\in\Psi_\alpha$,
\begin{equation}
\norm{e(t)}<\psi(t)\quad\text{for all } t\ge0,
\label{eq:objective}
\end{equation}
with all closed-loop signals bounded.

\section{Velocity Information from Delayed Positions}
\label{sec:dq}
The unmeasured velocity error $\dot e$ is replaced by the difference quotient
\begin{equation}
\hat v(t):=\frac{e(t)-e(t-h)}{h},
\label{eq:vhat}
\end{equation}
where, for $t<h$, the delayed value is taken from the rest history of Assumption~\ref{ass:sys}(e). Since $\dot q(0)=0$, $\dot e$ is continuous at $t=0$, and the relations below hold for all $t\ge0$. We discuss three properties of $\hat v$; only the first enters the analysis.

The difference quotient, $\hat v(t)=\frac1h\int_{t-h}^{t}\dot e(\theta)\dd\theta$, is the average velocity error over the last $h$ seconds. Since $\dot e(t)-\dot e(\theta)=\int_\theta^t\ddot e(r)\dd r$ and $\int_{t-h}^{t}(t-\theta)\dd\theta=h^2/2$,
\begin{equation}
\norm{\dot e(t)-\hat v(t)}\le\frac h2\operatorname*{ess\,sup}_{\theta\in[t-h,t]}\norm{\ddot e(\theta)}.
\label{eq:vbound}
\end{equation}

The transfer function from $\dot e$ to $\hat v$ is $(1-e^{-\sigma h})/(\sigma h)$~\cite{French2009,Selivanov2018}, to be compared with $1/(\varepsilon\sigma+1)$ for the dirty derivative~\cite{Kelly2005,Loria2016} and $1/(\varepsilon\sigma+1)^2$ for a high-gain observer with a double pole~\cite{KhalilPraly2014}. All three recover $\dot e$ as $h,\varepsilon\to0$ and delay slow signals by $h/2$, $\varepsilon$ and $2\varepsilon$, respectively. The latter two require one and two states per joint, whereas $\hat v$ has no state governed by a differential equation: it only needs the positions over one delay interval.

Differentiating~\eqref{eq:vhat} yields
\begin{equation}
\dot{\hat v}(t)=\frac1h\int_{t-h}^{t}\ddot e(\theta)\dd\theta ,
\label{eq:dDh}
\end{equation}
and $\ddot e$ contains $M(q)^{-1}\tau$ by~\eqref{eq:acc}. Hence a torque applied at time $t$ acts fully on the feedback signal only at time $t+h$. To see the consequence, let $n=1$, $M=\lambda$ constant, $N=d=0$, $\ddot q_d=0$ and $\tau=-\kappa\hat v$ with constant $\kappa$. Then $\ddot e=-(\kappa/\lambda)\hat v$, and~\eqref{eq:dDh} becomes
\begin{equation}
\dot{\hat v}(t)=-\frac{\kappa}{\lambda h}\int_{t-h}^{t}\hat v(\theta)\dd\theta ,
\label{eq:scalar}
\end{equation}
which is exponentially stable if and only if $\kappa h/\lambda<\pi^2/2$~\cite{Hale1993,Michiels2014}: $\sigma=\mathrm{i}\omega$ is a characteristic root only if $\omega h=\pm(2j+1)\pi$ and $\kappa h/\lambda=(\omega h)^2/2$, $j\in\mathbb{N}_0$, so that, at the boundary, an oscillation of period $2h$ persists. For $n>1$ and constant $M$, the same holds for $u^\top\hat v$ along each unit eigenvector $u$ of $M$, with $\lambda$ the corresponding eigenvalue. The delay must therefore be short compared with $\lambda/\kappa$. In funnel control, however, $\kappa$ grows as the error approaches the funnel boundary, so a fixed delay can be admissible only if the gain remains bounded.

\section{Main Result}
\label{sec:main}
Let $\psi\in\Psi_\alpha$ and consider the controller
\begin{subequations}\label{eq:ctrl}
\begin{align}
\sh(t) &= \hat v(t)+k\,e(t), \label{eq:ctrl_s}\\
\tau(t) &= -\kappa(t)\,\sh(t),\qquad \kappa(t)=\frac{\kappa_0}{1-\norm{\sh(t)}^2/b(t)^2}, \label{eq:ctrl_tau}
\end{align}
\end{subequations}
with $\hat v$ as in~\eqref{eq:vhat} and $b:=c_b\psi$, where $k>\alpha$, $c_b\in(0,k-\alpha)$, $\kappa_0>0$ and $h>0$ are design parameters. If $\hat v$ is replaced by $\dot e$, then~\eqref{eq:ctrl} is a full-state funnel controller for the surrogate $s:=\dot e+ke$, similar to~\cite{Dennstaedt2025low}; the difference quotient is the only modification.

\subsection{Closed loop with the finite-difference remainder}
Integration by parts gives the Taylor expansion of the delayed error with integral remainder,
\begin{equation}
e(t-h)=e(t)-h\,\dot e(t)+\int_{t-h}^{t}(\theta-t+h)\,\ddot e(\theta)\dd\theta ,
\label{eq:taylor}
\end{equation}
so that the difference quotient~\eqref{eq:vhat} reads $\hat v=\dot e-\delta$ with the remainder
\begin{equation}
\delta(t):=\frac1h\int_{t-h}^{t}(\theta-t+h)\,\ddot e(\theta)\dd\theta .
\label{eq:delta}
\end{equation}
Hence $\sh=s-\delta$ with $s:=\dot e+ke$, and, by~\eqref{eq:acc} and~\eqref{eq:ctrl}, the closed-loop system reads, in the coordinates $(e,s)$,
\begin{subequations}\label{eq:cl}
\begin{align}
\dot e&=-ke+s,\label{eq:cl_e}\\
\dot s&=M(q)^{-1}\big(-\kappa(t)(s-\delta)-N(q,\dot q)-d\big)-\ddot q_d+k\dot e.\label{eq:cl_s}
\end{align}
\end{subequations}
For $\delta=0$, \eqref{eq:cl} is the full-state funnel loop; the delay enters only through the remainder $\delta$. Unlike in artificial-delay designs~\cite{FridmanSelivanov2025}, the delay is thus not part of the stabilizing mechanism but an error to be kept small, which makes the resulting bound conservative. By~\eqref{eq:delta}, $\delta$ is small if the acceleration is bounded, which holds if the gain is bounded, that is, if $\sh$ stays away from $b$. The analysis breaks this circle with nested tubes.

\subsection{Main result}
Fix analysis parameters $\rho\in(0,1)$, $\beta\in(0,\min\{\rho,1-\rho\})$ and $\eb\in[\rho c_b/(k-\alpha),1)$, and consider the tubes
\begin{equation}
\norm{e}\le\eb\,\psi,\qquad \norm{s}\le\rho\, b,\qquad \norm{\sh}\le(\rho+\beta)\, b .
\label{eq:apriori}
\end{equation}
In these tubes, $\norm{\dot e}\le k\norm e+\norm s\le(k\eb+\rho c_b)\psi$, so the forces and the input-free terms in~\eqref{eq:cl_s} are bounded by
\begin{align*}
W_N(t)&:=\max\big\{\norm{N(x,v)}:\ \norm x\le\eb\psi(t)+\|q_d\|_\infty,\\
&\qquad\ \norm v\le(k\eb+\rho c_b)\psi(t)+\|\dot q_d\|_\infty\big\}+\|d\|_\infty,\\
W_q(t)&:=\|\ddot q_d\|_\infty+k(k\eb+\rho c_b)\psi(t),
\end{align*}
and the gain is at most $\kb:=\kappa_0/(1-(\rho+\beta)^2)$. Let
\begin{align}
R&:=\sup_{t\ge0}\frac1{b(t)}\Big(\frac{W_N(t)}{\sqrt{\lm\lM}}+W_q(t)\Big)+\alpha\rho,\label{eq:R}\\
\eta&:=\sup_{t\ge0}\frac1{b(t)}\Big(\frac{W_N(t)}{\lm}+\|\ddot q_d\|_\infty\Big),\label{eq:eta}
\end{align}
and define $h^*>0$ by
\begin{equation}
h^*e^{\alpha h^*}=\frac{2\beta}{\kb(\rho+\beta)/\lm+\eta}.
\label{eq:hstar}
\end{equation}
Here, $R$ is the rate, relative to the funnel width, at which the forces and the reference push $s$ outward, and $\eta\,b$ bounds the acceleration caused by the forces and the reference.

On the boundary $\norm s=\rho b$, the forces and the reference push $s$ outward at a rate of at most $R$, relative to the funnel width. Since $\norm\delta\le\beta b$ for $h\le h^*$, the gain there is at least $\kappa_0/(1-(\rho-\beta)^2)$, and the controller pushes $s$ inward at a rate of at least
\begin{equation}
L(\rho):=\frac{\kappa_0}{1-(\rho-\beta)^2}\Big(\frac{\rho}{\lM}-\frac{\beta}{\sqrt{\lm\lM}}\Big).
\label{eq:L}
\end{equation}
We require the inward rate to dominate:
\begin{equation}
L(\rho)>R.
\label{eq:cond}
\end{equation}
The following theorem shows that, under this condition, every delay $h\in(0,h^*]$ keeps the tracking error within the funnel and all signals bounded.

\begin{theorem}\label{thm:main}
Consider system~\eqref{eq:EL} and controller~\eqref{eq:ctrl} under Assumption~\ref{ass:sys}. Let $\rho$, $\beta$ and $\eb$ be as above and satisfy~\eqref{eq:cond}, and suppose that $\norm{e(0)}\le\eb\,\psi(0)$ and $\norm{k\,e(0)-\dot q_d(0)}<\rho\,b(0)$. Then, for every $h\in(0,h^*]$, the closed-loop system has a unique global solution and, for all $t\ge0$,
\begin{enumerate}
  \item[(i)] $\norm{e(t)}\le\eb\,\psi(t)<\psi(t)$;
  \item[(ii)] $\norm{\dot e(t)+k\,e(t)}<\rho\,b(t)$;
  \item[(iii)] $\kappa_0\le\kappa(t)<\kb$ and $\norm{\tau(t)}<\kb(\rho+\beta)\,b(t)$.
\end{enumerate}
\end{theorem}

\begin{proof}
By Assumption~\ref{ass:sys}(e), $s(0)=ke(0)-\dot q_d(0)$, so $\norm{s(0)}<\rho b(0)$. Moreover, $\ddot e=-\ddot q_d$ on $[-h,0)$, and
\eqref{eq:delta}, \eqref{eq:eta} and \eqref{eq:hstar} give
\[
\norm{\delta(0)}
\le \frac h2\|\ddot q_d\|_\infty
\le \beta b(0).
\]
Thus $\norm{\sh(0)}<(\rho+\beta)b(0)<b(0)$, and the
controller is initially well defined. On the first delay
interval, the delayed position equals $q^0$, and the
closed-loop dynamics form a Carath\'eodory ordinary
differential equation in $(q,\dot q)$, locally Lipschitz
in the state wherever $\norm{\sh}<b$. The method of
steps~\cite{Hale1993} therefore yields a unique maximal
solution on $[0,\omega)$. Since $\dot q(0)=0$ agrees with
the rest history, $s$, $\sh$ and $\delta$ are continuous.

Suppose that, at some first time $t_1<\omega$,
\[
\norm{s(t_1)}=\rho b(t_1)
\quad\text{or}\quad
\norm{\sh(t_1)}=(\rho+\beta)b(t_1).
\]
Both corresponding non-strict bounds then hold on
$[0,t_1]$. Since
$\psi(\theta)\le e^{\alpha(t-\theta)}\psi(t)$ for
$0\le\theta\le t$, variation of constants
in~\eqref{eq:cl_e} gives, with $\mu:=e^{-(k-\alpha)t}$,
\begin{equation}
\frac{\norm{e(t)}}{\psi(t)}
\le
\mu\,\frac{\norm{e(0)}}{\psi(0)}
+(1-\mu)\frac{\rho c_b}{k-\alpha}
\le \eb .
\label{eq:ebound}
\end{equation}
Hence all bounds in~\eqref{eq:apriori} hold on $[0,t_1]$.
In particular, $\norm{N+d}\le W_N$,
$\kappa\le\kb$ and
$\norm{\tau}\le\kb(\rho+\beta)b$. It follows
from~\eqref{eq:acc} and \eqref{eq:eta} that
\begin{equation}
\norm{\ddot e(t)}
\le
\left(\frac{\kb(\rho+\beta)}{\lm}+\eta\right)b(t)
\label{eq:accbound}
\end{equation}
almost everywhere on $[0,t_1]$, while
$\norm{\ddot e}\le\eta b(0)$ on $[-h,0)$.
Extend $b$ by $b(\theta)=b(0)$ for $\theta<0$.
Then $b(\theta)\le e^{\alpha h}b(t)$ whenever
$\theta\in[t-h,t]$. Substituting these estimates
into~\eqref{eq:delta} yields
\begin{equation}
\norm{\delta(t)}
\le
\frac h2 e^{\alpha h}
\left(\frac{\kb(\rho+\beta)}{\lm}+\eta\right)b(t)
\le \beta b(t),
\qquad t\in[0,t_1],
\label{eq:dbound}
\end{equation}
where the last inequality follows from~\eqref{eq:hstar}
and $h\le h^*$.

To exclude contact with $\norm{s}=\rho b$, let
$V:=\norm{s}^2/b^2$. By continuity of $L$ and~\eqref{eq:cond},
choose $\rho'\in(\max\{\beta,\sqrt{V(0)}\},\rho)$ such that
$L(\nu)>R$ for all $\nu\in[\rho',\rho]$.
Setting $A:=M(q)^{-1}$ and $x:=\norm{A^{1/2}s}$,
the inertia bounds give
\[
x\ge x_0:=\frac{\norm{s}}{\sqrt{\lM}},
\qquad
|s^\top Az|\le\frac{x\norm{z}}{\sqrt{\lm}}
\quad\text{for all }z\in\mathbb{R}^n.
\]
Using~\eqref{eq:cl_s}, $\dot b\ge-\alpha b$ and
$\norm{k\dot e-\ddot q_d}\le W_q$, we obtain
almost everywhere
\begin{align}
\frac{b^2}{2}\dot V
&=-\kappa s^\top As+\kappa s^\top A\delta
  -s^\top A(N+d) \notag\\
&\quad+s^\top(k\dot e-\ddot q_d)
  -\frac{\dot b}{b}\norm{s}^2 \notag\\
&\le -x(\kappa x-B)
  +(W_q+\alpha\norm{s})\norm{s},
\label{eq:Vs}
\end{align}
where $B:=(\kappa\norm{\delta}+W_N)/\sqrt{\lm}$.

Whenever $\nu:=\norm{s}/b\in[\rho',\rho]$,
\eqref{eq:dbound} gives
$\norm{\sh}\ge(\nu-\beta)b$, and hence
\[
\kappa\ge\frac{\kappa_0}{1-(\nu-\beta)^2}.
\]
Since $L(\nu)>R\ge0$, the factor
$\nu/\lM-\beta/\sqrt{\lm\lM}$ is positive, so
\[
\kappa\left(
\frac{\nu}{\lM}-\frac{\beta}{\sqrt{\lm\lM}}
\right)\ge L(\nu)>R.
\]
Together with~\eqref{eq:R} and \eqref{eq:dbound},
this implies $\kappa x_0>B$. Consequently,
$-x(\kappa x-B)\le-x_0(\kappa x_0-B)$ for $x\ge x_0$.
Dividing~\eqref{eq:Vs} by $\norm{s}b$ therefore yields
\begin{align}
\frac{b}{2\norm{s}}\dot V
&\le
-\kappa\left(
\frac{\nu}{\lM}-\frac{\beta}{\sqrt{\lm\lM}}
\right)
+\frac{W_N}{b\sqrt{\lm\lM}}
+\frac{W_q}{b}+\alpha\nu \notag\\
&\le -L(\nu)+R<0.
\label{eq:VL}
\end{align}

If $\norm{s(t_1)}=\rho b(t_1)$, continuity and
$V(0)<\rho'^2$ provide a last time $t_0<t_1$ with
$V(t_0)=\rho'^2$. Then
$\rho'^2\le V(t)\le\rho^2$ on $[t_0,t_1]$.
Since $V$ is absolutely continuous,
integrating~\eqref{eq:VL} gives
$V(t_1)<V(t_0)=\rho'^2$, a contradiction.
Thus $\norm{s(t_1)}<\rho b(t_1)$, and
\eqref{eq:dbound} further gives
\[
\norm{\sh(t_1)}
\le\norm{s(t_1)}+\norm{\delta(t_1)}
<(\rho+\beta)b(t_1),
\]
contradicting the definition of $t_1$.

The bounds therefore hold throughout $[0,\omega)$.
Since $\psi$, $q_d$ and $\dot q_d$ are bounded,
\eqref{eq:ebound} and $\dot e=s-ke$ imply boundedness
of $(q,\dot q)$. Moreover,
$\norm{\sh}/b<\rho+\beta<1$ and $\inf_{t\ge0}b(t)>0$,
so the solution remains uniformly separated from the
singularity of the controller. The continuation theorem
then gives $\omega=\infty$. Finally,
\eqref{eq:ebound}, $\norm{s}<\rho b$ and
$\norm{\sh}<(\rho+\beta)b$, together
with~\eqref{eq:ctrl}, establish (i)--(iii).
\end{proof}

Condition~\eqref{eq:cond} does not restrict $\kappa_0$: as $\rho\to1$ and $\beta\to0$, $L(\rho)$ grows without bound, whereas $R$ remains bounded. Hence, for every $\kappa_0>0$ and every initial error with $\norm{e(0)}<\psi(0)$ and $\norm{ke(0)-\dot q_d(0)}<b(0)$, there exist $\rho$, $\beta$, $\eb$ and an admissible delay $h^*>0$, and, as in full-state funnel control and unlike in~\cite{Lanza2024derivfree}, no gain threshold has to be known. The gain adapts, but remains below $\kb$, and it is this a priori bound that renders a fixed delay admissible.

\begin{figure}[t]
\centering
\includegraphics[width=\textwidth]{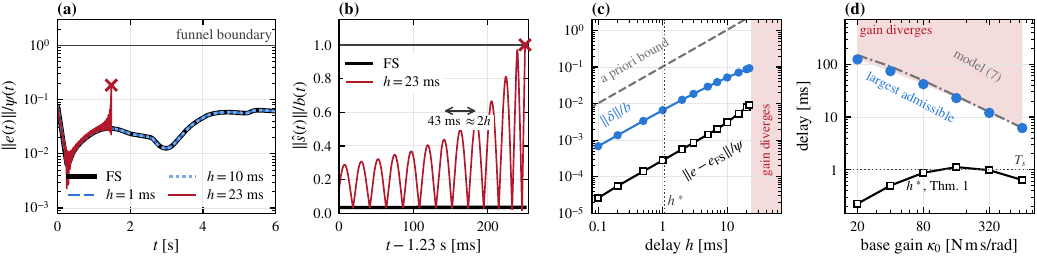}
\caption{Continuous-time loop, $\kappa_0=160$\,N\,m\,s/rad unless stated. (a)~Normalized error for full-state (FS) funnel control and three delays; $\times$:~gain diverges. (b)~Normalized surrogate before the divergence, $h=23$\,ms. (c)~Largest $\norm\delta$ and deviation from FS over $6$\,s; dashed: bound~\eqref{eq:dbound}. (d)~Largest admissible delay, prediction of~\eqref{eq:scalar} ($\lambda=\lm$, largest FS gain) and $h^*$ ($\rho$, $\beta$ optimized); dotted: $1$\,ms.}
\label{fig:theory}
\end{figure}
\begin{remark}\label{rem:size}
By~\eqref{eq:hstar}, $\kb h^*/\lm<2\beta/(\rho+\beta)<1$, so $h^*$ lies inside the stability region $\kappa h/\lambda<\pi^2/2$ of the frozen model~\eqref{eq:scalar} at the largest gain and the smallest inertia. Moreover, \eqref{eq:R} and~\eqref{eq:cond} imply $\kb>R\lM/\rho\ge W_N\sqrt{\lM/\lm}/(\rho b)$, and hence
\[
h^*<\inf_{t\ge0}\frac{\lm\,b(t)}{W_N(t)}\sqrt{\frac{\lm}{\lM}} .
\]
The delay must thus be short compared with the time $\lm b/W_N$ in which the largest force changes the joint velocity by the funnel width. Geared drives, whose reflected rotor inertia dominates $M(q)$, have a larger $\lm$ and a smaller ratio $\lM/\lm$ than direct drives, and therefore admit longer delays if gravity and load torques, rather than reflected friction, dominate $W_N$; the inertia-independent terms $W_q/b+\alpha\rho$ in $R$ eventually limit this benefit. Although the control law uses neither a model nor derivatives of the reference, $h^*$ depends on bounds on both, as do the delay in~\cite{IlchmannSangwin2004,French2009} and the sampling time in~\cite{LanzaSD2024,Lanza2024derivfree}; $\rho$, $\beta$ and $\kappa_0$ can be chosen to maximize $h^*$. Theorem~\ref{thm:main} assumes continuous measurements; with zero-order hold and $h=T_s$, \eqref{eq:ctrl} becomes the backward difference of~\cite{Lanza2024derivfree}, and extending the guarantee requires bounding the inter-sample drift, for which the a~priori bound $\kappa\le\kb$ is the key ingredient, as in~\cite{LanzaSD2024}.
\end{remark}
\begin{remark}\label{rem:ic}
The admissible choice $\eb=\max\{\norm{e(0)}/\psi(0),\rho c_b/(k-\alpha)\}$ is smaller than one if $\norm{e(0)}<\psi(0)$. For $\dot q_d(0)=0$, the stronger condition $\norm{k e(0)-\dot q_d(0)}<\rho b(0)$ reads $\norm{e(0)}/\psi(0)<\rho c_b/k$, which is met by selecting $\psi(0)$ sufficiently large. Independently of the delay, the controller is defined at $t=0$ only if $\norm{e(0)}/\psi(0)<c_b/k$, since $\sh(0)\approx k\,e(0)$ at rest; with $c_b<k-\alpha$, this ratio stays below $1-\alpha/k$. The rest start in Assumption~\ref{ass:sys}(e) can be relaxed: the proof only needs $\dot e$ continuous at $t=0$ and $\norm{\ddot e}\le\eta b(0)$ on $[-h,0)$.
\end{remark}

\section{Simulation Study}
\label{sec:sim}
Consider a two-link manipulator in the vertical plane with uniform links of lengths $0.30$ and $0.25$\,m and masses $2$ and $1$\,kg, and a payload $m_p\in[0,1]$\,kg, nominally $0.5$\,kg, at the tip. The joints are geared: the reflected rotor inertias $1.2$ and $0.8$\,kg\,m$^2$ (gear ratio $160$) enter the diagonal of $M(q)$. The viscous and Coulomb friction coefficients are $1.0$ and $0.6$\,N\,m\,s/rad and $1.0$ and $0.6$\,N\,m, the latter smoothed by $\tanh(\dot q_i/0.05)$. Over the payload range, Assumption~\ref{ass:sys}(a) holds with $\lm=0.815$ and $\lM=1.79$\,kg\,m$^2$. Unless stated otherwise, $d(t)=(0.5\sin2t,\ 0.2\cos3t)$\,N\,m, $q_d(t)=\big(0.5(1-\cos1.2t),\ 0.3+0.4(1-\cos1.6t)\big)$\,rad and $e(0)=(0.1,-0.05)$\,rad, and the controller~\eqref{eq:ctrl} uses $\psi(t)=1.4e^{-2t}+0.1$\,rad, $k=10$\,s$^{-1}$, $c_b=7.7$\,s$^{-1}$ and $\kappa_0=160$\,N\,m\,s/rad. The controller runs with zero-order hold at $100$\,kHz to emulate continuous time (Sections~\ref{sec:nf}, \ref{sec:ic}) and at $1$\,kHz with an 18-bit joint-side encoder and $\pm10\,\mu$rad noise (Section~\ref{sec:cmp}).

With $W_N$ evaluated on a grid over configurations and payloads ($1\%$ margin), $\rho=0.48$, $\beta=0.11$ and $\eb=\rho c_b/(k-\alpha)=0.462$, one obtains $L(\rho)=32.8$\,s$^{-1}>R=31.1$\,s$^{-1}$, $\kb=245$\,N\,m\,s/rad and $\eta=27.9$\,s$^{-1}$, and~\eqref{eq:hstar} gives $h^*=1.07$\,ms. Since $\norm{e(0)}=0.11\le\eb\psi(0)=0.69$\,rad and $\norm{ke(0)-\dot q_d(0)}=1.12<\rho b(0)=5.54$\,rad/s, Theorem~\ref{thm:main} certifies $h=1$\,ms, one sampling period of a $1$\,kHz joint controller.

\subsection{Effect of delay}
\label{sec:nf}
Fig.~\ref{fig:theory} compares the continuous-time loop with full-state (FS) funnel control, that is,~\eqref{eq:ctrl} with $\dot e$ in place of $\hat v$. In Fig.~\ref{fig:theory}(a), for the certified delay $h=1$\,ms and for $h=10$\,ms, the error cannot be distinguished from that of FS funnel control, whereas for $h=23$\,ms the gain diverges after $1.48$\,s. Fig.~\ref{fig:theory}(b) shows the mechanism predicted by~\eqref{eq:scalar}: an oscillation of period $43$\,ms${}\approx2h$ grows until $\sh$ reaches the boundary. In Fig.~\ref{fig:theory}(c), $\delta$ and the deviation from FS grow linearly with $h$; at $h=1$\,ms, $\norm\delta$ is $16$ times below~\eqref{eq:dbound} and the deviation is $3\cdot10^{-4}\psi$, and even at $h=22$\,ms it stays below $1\%$ of the funnel radius. Fig.~\ref{fig:theory}(d) shows that the largest admissible delay is inversely proportional to $\kappa_0$ and that~\eqref{eq:scalar} predicts it to within $1$ to $23\%$. The certified delay is $10$ to $540$ times shorter, mainly because~\eqref{eq:hstar} contains $2\beta/(\rho+\beta)<1$ where~\eqref{eq:scalar} allows $\pi^2/2$, and because $\kb$ is a worst-case gain. Still, $h^*$ is close to one sampling period at $1$\,kHz for $\kappa_0$ between $80$ and $320$\,N\,m\,s/rad, so that the certificate can serve as a design rule.

\begin{figure}[t]
\centering
\includegraphics[width=0.75\linewidth]{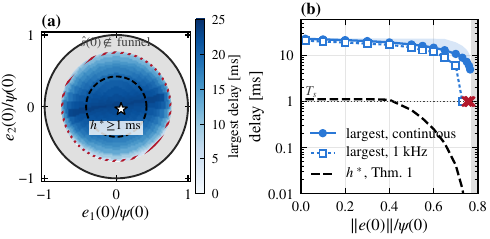}
\caption{Initial errors, $\kappa_0=160$\,N\,m\,s/rad. (a)~Largest admissible delay in continuous time; $\star$: nominal $e(0)$; dashed circle: $h^*\ge1$\,ms; grey: $\sh(0)$ outside the funnel; hatched: fails at $1$\,kHz for $h=T_s$. (b)~Worst direction (band: all) in continuous time and at $1$\,kHz ($h=NT_s$; $\times$: fails for $h=T_s$), and $h^*$.}
\label{fig:ic}
\end{figure}

\subsection{Initial errors}
\label{sec:ic}
Fig.~\ref{fig:ic}(a) maps the largest admissible delay over all initial errors for which the controller is defined. Over most of this region, about $20$\,ms is admissible; only near its edge does the direction matter, with errors in joint~2 being the worst case ($5$\,ms). In Fig.~\ref{fig:ic}(b), $h^*\ge1$\,ms up to $\norm{e(0)}=0.42\psi(0)$, beyond which $h^*$ falls much faster than the true limit, because the initial condition of Theorem~\ref{thm:main} forces $\rho\to1$ and $\beta\to0$. At $1$\,kHz, delays of several samples remain admissible up to $0.7\psi(0)$, but from $0.75\psi(0)$ on the loop fails even for $h=T_s$, also with velocity measurements, because the initial gain of about $20\kappa_0$ is held over a full sampling period. Near the edge, the initial torque reaches $143$\,kN\,m ($0.765\psi(0)$), so $\psi(0)$ should be chosen generously; combining the difference quotient with funnel control under input saturation~\cite{Hopfe2010} is left for future work. The bounded-input SH controller (Section~\ref{sec:cmp}) keeps the funnel for all these initial errors with at most $167$\,N\,m.

\begin{figure}[t]
\centering
\includegraphics[width=\textwidth]{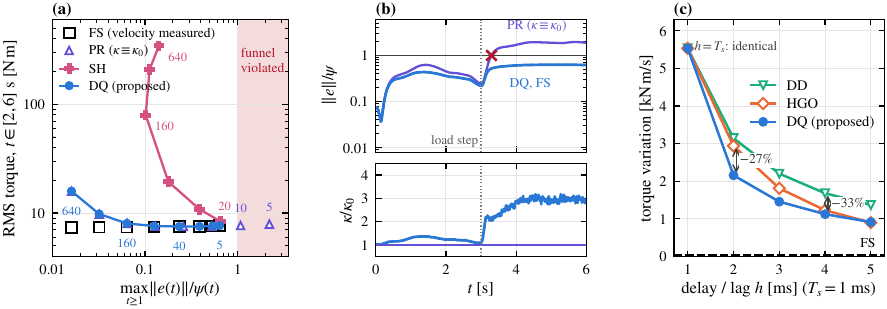}
\caption{Comparison at $1$\,kHz with the 18-bit encoder (methods: Section~\ref{sec:cmp}). (a)~Worst normalized error for $t\ge1$\,s versus RMS torque for $t\in[2,6]$\,s as the gain is varied (labels: $\kappa_0$ in N\,m\,s/rad or $\beta$ in N\,m). (b)~Ablation of the funnel gain: $\kappa_0=10$\,N\,m\,s/rad, payload $1$\,kg, load step $(6,2)$\,N\,m at $t=3$\,s. (c)~Torque variation per second versus the delay or lag, mean of five noise realizations.}
\label{fig:cmp}
\end{figure}

\begin{table}[t]
\centering
\caption{Comparison at $1$\,kHz with the 18-bit encoder: structure and $100$ randomized scenarios (nominal tuning)}
\label{tab:mc}
\small\setlength{\tabcolsep}{5pt}
\begin{tabular}{@{}lccccccc@{}}
\toprule
 & FS & \multicolumn{2}{c}{\textbf{DQ}} & HGO & DD & PR & SH \\
\cmidrule(lr){3-4}
Delay/lag $h$ [ms] & -- & 1 & 2 & 2 & 2 & 1 & 1 \\
States/joint & -- & 0$^{a}$ & 0$^{a}$ & 2 & 1 & 0$^{a}$ & 0$^{a}$ \\
Gain adapts & yes & yes & yes & yes & yes & no & -- \\
Guarantee & Thm.~\ref{thm:main}$^{b}$ & Thm.~\ref{thm:main}$^{c}$ & --$^{d}$ & --$^{e}$ & --$^{f}$ & -- & \cite{Lanza2024derivfree}$^{g}$ \\
\midrule
Funnel kept [\%] & 100 & 100 & 100 & 100 & 100 & 100 & 100 \\
Worst $\norm{e}/\psi$, $t\ge1$ & .094 & .094 & .094 & .094 & .094 & .095 & .318 \\
RMS $\norm\tau$ [N\,m]$^{h}$ & 7.41 & 7.99 & 7.57 & 7.60 & 7.62 & 8.00 & 60.2 \\
Variation [kN\,m/s]$^{h}$ & 0.04 & 5.48 & 2.16 & 2.95 & 3.15 & 5.44 & 93.0 \\
Peak $\norm\tau$ [N\,m] & 242 & 242 & 242 & 242 & 242 & 238 & 167 \\
Largest $T_s$ [ms] & 10 & 9 & -- & -- & -- & 9 & 6 \\
\bottomrule
\end{tabular}

\vspace{2pt}
\parbox{0.9\linewidth}{\footnotesize $^{a}$\,stores $h/T_s$ past samples. $^{b}$\,proof with $\delta=0$; cf.~\cite{Dennstaedt2025low}. $^{c}$\,in continuous time. $^{d}$\,$h>h^*$. $^{e}$\,for $\varepsilon$ below a threshold that is not explicit~\cite{Chowdhury2019,Dimanidis2020}. $^{f}$\,asymptotic tracking only~\cite{Loria2016}. $^{g}$\,needs $e(0)=0$ and, here, $\beta\gtrsim900$\,N\,m and $T_s<0.05$\,ms~\cite[Thm.~4.1]{Lanza2024derivfree}. $^{h}$\,median over the scenarios, for $t\in[2,6]$\,s; variation: total variation of $\tau$ per second.}
\end{table}

\subsection{Comparison results}
\label{sec:cmp}
All controllers use only sampled positions and $q_d$, except the benchmark FS, which measures the velocity. DQ is the proposed controller with $h=T_s$. The high-gain observer (HGO)~\cite{KhalilPraly2014} and the dirty derivative (DD)~\cite{Loria2016} replace $\hat v$ in~\eqref{eq:ctrl_s}, implemented as two and one low-pass sections driven by the backward difference of the measured error, with poles $(h-T_s)/(h+3T_s)$ and $(h-T_s)/(h+T_s)$, so that, like the difference quotient (DQ), they reproduce ramps and delay slow signals by $h/2$; for $h=T_s$, both coincide with DQ. Proportional-retarded (PR) control~\cite{RamirezNeria2019} is DQ with $\kappa\equiv\kappa_0$, the ablation of the funnel gain. The sample-and-hold (SH) controller of~\cite{Lanza2024derivfree} uses the same two samples in a normalized law with gain $\beta$ and threshold $\lambda=0.9$; since its conditions cannot be met at $1$\,kHz (Table~\ref{tab:mc}), the most accurate value $\beta=150$\,N\,m is used. Input filters~\cite{Dennstaedt2025filter,Schaa2026} require a constant input matrix and are not included.

Over its whole gain range (Fig.~\ref{fig:cmp}(a)), DQ is exactly as accurate as FS for every $\kappa_0$ between $5$ and $640$\,N\,m\,s/rad, at RMS torques of $7.5$ to $16$\,N\,m (FS: $7.4$\,N\,m). SH keeps the funnel for $\beta\ge20$\,N\,m, but its error does not fall below $0.10\psi$, which takes ten times the RMS torque of DQ. PR matches DQ for $\kappa_0\ge80$\,N\,m\,s/rad and leaves the funnel for $\kappa_0\le10$\,N\,m\,s/rad. In Fig.~\ref{fig:cmp}(b), at $\kappa_0=10$\,N\,m\,s/rad with the heavier payload and a load step, PR leaves the funnel, whereas DQ raises its gain to about $3\kappa_0$ and keeps the error at $0.63\psi$, exactly as FS; the delay does not impair the gain adaptation.

At equal lag, the three estimators yield the same tracking error, so they differ only in how encoder noise reaches the torque (Fig.~\ref{fig:cmp}(c)). For two to four sampling periods, DQ causes the least torque variation, $27\%$ less than HGO at $h=2$\,ms; HGO draws level only at $h=5$\,ms, and DQ stays $31$ to $34\%$ below DD at every lag. DQ thus attains the noise attenuation of an observer without a state governed by a differential equation, and with an explicit admissible delay.

Table~\ref{tab:mc} evaluates the nominal tunings in $100$ scenarios with random payload, initial error ($\le0.15$\,rad), disturbance amplitudes and load steps (up to $4$ and $1.5$\,N\,m), which exceed the bounds used for $h^*$. All controllers keep the funnel in every scenario, and all funnel controllers reach the same worst error. SH has a three-times-larger error and a $7.5$-times-larger RMS torque, but the lowest peak torque. With the nominal gains, DQ tolerates sampling periods up to $9$\,ms, SH up to $6$\,ms.

\section{Conclusions}
We have shown that the velocity error in a funnel controller for fully actuated Euler--Lagrange systems can be replaced by a difference quotient of the position error over an artificial delay, while retaining the classical unbounded funnel gain. Despite the delayed feedback, the boundedness of the gain along closed-loop trajectories ensures that a sufficiently small fixed delay can be accommodated. The resulting controller requires neither an observer nor additional filter states, and relies solely on position measurements, without requiring a system model or derivatives of the reference trajectory. The analysis provides a continuous-time guarantee, while numerical studies demonstrate close agreement with full-state funnel control and favorable performance under encoder quantization compared with other position-only implementations. Future work will address less conservative delay conditions, potentially using Lyapunov--Krasovskii techniques~\cite{FridmanSelivanov2025}, and extend the theoretical guarantees to zero-order-hold implementation and encoder quantization. Experimental validation on a robotic manipulator will also be pursued.

\end{document}